\documentclass[conference]{IEEEtran}
\pdfoutput=1
\usepackage[cmex10]{amsmath}
\usepackage{amssymb,amsthm,mathrsfs}
\usepackage{graphicx}
\usepackage{pgfplots}
\usepackage{tikz}
 \usetikzlibrary{decorations}
 \usetikzlibrary{arrows}
 \usetikzlibrary{plotmarks}
\usepackage{subfigure}
\usepackage{multirow}

\newtheorem{theorem}{Theorem}

\newtheorem{corollary}[theorem]{Corollary}
\theoremstyle{definition}

\theoremstyle{remark}
    
    \newenvironment{sketch}{\par\noindent{\it Sketch of Proof:}\quad}{\hfill\qed\par\smallskip}

\DeclareMathOperator*{\tr}{tr}
\DeclareMathOperator*{\rank}{rank}
\DeclareMathOperator*{\E}{E}

\begin{document}
\sloppy
\IEEEoverridecommandlockouts
\title{An Upper Bound on the Capacity of \\Vector Dirty Paper with Unknown Spin and Stretch}
\author{
\IEEEauthorblockN{David T.H. Kao}
 \IEEEauthorblockA{Cornell University -- Ithaca, NY \\
  \texttt{davidkao@cornell.edu}
 }
 \and
 \IEEEauthorblockN{Ashutosh Sabharwal}
 \IEEEauthorblockA{Rice University -- Houston, TX \\
  \texttt{ashu@rice.edu}
 }
 \thanks{This work has been supported in part by NSF CNS award 1012921, and was completed while David T.H. Kao was at Rice University.}
}

\maketitle

\begin{abstract}

Dirty paper codes are a powerful tool for combating known interference. However, there is a significant difference between knowing the transmitted interference sequence and knowing the received interference sequence, especially when the channel modifying the interference is uncertain.

We present an upper bound on the capacity of a compound vector dirty paper channel where although an additive Gaussian sequence is known to the transmitter, the channel matrix between the interferer and receiver is uncertain but known to lie within a bounded set. Our bound is tighter than previous bounds in the low-SIR regime for the scalar version of the compound dirty paper channel and employs a construction that focuses on the relationship between the dimension of the message-bearing signal and the dimension of the additive state sequence. Additionally, a bound on the high-SNR behavior of the system is established.

\end{abstract}

\section{Introduction}

The benefit of transmitter side information has been studied in many forms. The case of causal state information for the discrete memoryless channel was first studied in \cite{Shannon:1958}, and noncausal state information was subsequently considered in \cite{GP:1980,HE:83}. The noncausal case was later specialized to the point-to-point AWGN channel with additive Gaussian state in \cite{Costa:1983}, the so-called dirty paper channel, wherein it was shown that, surprisingly, interference perfectly known at the transmitter may be completely mitigated through a clever binning scheme. 

This dirty paper coding (DPC) approach is especially applicable to certain scenarios in multiuser wireless communications: In the vector Gaussian broadcast channel, DPC enables a capacity achieving encoding scheme~\cite{CS:2003,WSS:2006} which bins each successive receiver's message against interference from messages intended for preceding receivers. In the cognitive interference channel, DPC is a useful tool for exploiting cognitive knowledge to mitigate interference~\cite{DMT:2006,RTD:2012}. 

A known limitation of DPC is its reliance on exact knowledge of channel state. Unfortunately, in large and distributed wireless networks, providing transmitters with full channel state knowledge incurs high overhead. Hence, often in practice, channel state is known with some uncertainty. In this paper we study how such channel state uncertainty inhibits the usefulness of transmitter side information. 
Specifically, we study a compound vector dirty paper channel, where side information about an additive vector Gaussian interference sequence is provided to the transmitter. Channel uncertainty is modeled by a set of possible channel matrices that transform the interference sequence prior to reception. The set studied contains all matrices with singular values less than a parameter $a_\mathsf{max}$, representing a known maximum amplification. The transmitter, unaware of the exact channel state, must reliably convey a message to the receiver.

Our compound formulation captures the subtle but important distinction between noncausal knowledge of transmitted interference and noncausal knowledge of received interference. With our model, we may better understand, for example, cognitive interference channels where the interference channel gains are unmeasured~\cite{GS:2007:dyspan}, as well as MIMO broadcast where elements of the channel matrix are unknown~\cite{KS:2012:asilomar}.

Compound versions of transmitter side information channels have been examined previously, with the most general formulation being \cite{MDT:2006}. In the work of \cite{KELW:2007}, a more precise model with a finite number of compound states was studied and an approach termed ``carbon copying'' was defined. Further extensions of \cite{KELW:2007} to specific Gaussian channels may be found in \cite{GS:2007:dyspan} where phase was uncertain, and \cite{ZKL:2007} where the message-bearing and interfering signals scaled proportionally. 

The main result of this paper is an upper bound on capacity of dirty paper channels when the interference signal may have both dimension greater than one and the potential to undergo a wide range of amplification. For scalar dirty paper, our bound is tighter than known bounds when maximum potential amplification of interference is high, i.e., the low signal-to-interference ratio (${\rm SIR}$) regime. For vector dirty paper in the low-${\rm SIR}$ regime, we find that uncertainty incurs an approximate prelog loss in capacity. When the unknown amplification is unbounded, the loss is exact and signifies a prelog capacity loss \emph{at all finite signal to noise ratios (${\rm SNR}$)}. A degrees of freedom ($\mathsf{DOF}$) upper bound also results, thus providing insight into the high-${\rm SNR}$ behavior of the system.

Additionally, to our knowledge, this work represents the first treatment of compound channels for \emph{vector} dirty paper, and the focus on vector channels reveals a relationship between the dimension of the interference, the dimension of the message bearing signal, and the resulting upper bound on $\mathsf{DOF}$. 

The paper is structured as follows. After defining our model in Section~\ref{sec:model}, we state and prove our upper bound in its most general form in Section~\ref{sec:result}. Section~\ref{sec:rankone} presents the bound applied to the more concrete cases of MISO and SIMO channels, and includes a comparison to bounds previously applied to the scalar compound dirty paper channel. We comment on high-${\rm SNR}$ behavior of the system in Section~\ref{sec:highSNR}, and summarize in Section~\ref{sec:remarks}.

\section{Problem Statement}
\label{sec:model}

\begin{figure}[tb]
\centering
\begin{tikzpicture}[yscale=0.9]
    \node (X) at (0,0)[left] {$\mathbf{x}$};
    \node (S) at (2,1.5)[above] {$\mathbf{s}$};
    \node (a) at (2.5,0.75)[right] {$\mathbf{A} \in \mathcal{A}$};
    \node (h) at (1,-0.75)[below] {$\mathbf{H}$};
    \node (timex) at (1,0) [draw,circle,inner sep=0pt]{$\times$};
    \node (times) at (2,0.75) [draw,circle,inner sep=0pt]{$\times$};
    \node (plus) at (2,0) [draw,circle,inner sep=0pt]{$+$};
    \node (Z) at (2,-0.75)[below] {$\mathbf{z}$};
    \node (Y) at (3,0)[right] {$\mathbf{y}$};
    \draw[very thick,-latex] (X)--(timex);
    \draw[very thick,-latex] (h)--(timex);
    \draw[very thick,-latex] (timex)--(plus);
    \draw[very thick,-latex] (S)--(times);
    \draw[very thick,-latex] (a)--(times);
    \draw[very thick,-latex] (times) -- (plus);
    \draw[very thick,-latex] (Z) -- (plus);
    \draw[very thick,-latex] (plus) -- (Y);
    \draw[very thick,dotted,-latex] (S) -| (X);
\end{tikzpicture}
\caption{Channel Model. An interference sequence is known to the transmitter, however the linear transformation of the interference prior to reception is known only to lie within some set.\vspace{-18pt}}\label{fig:chan}
\end{figure}
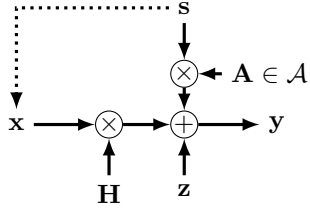

We consider the channel depicted by Figure~\ref{fig:chan} with input-output relationship characterized by
\begin{align}
    \mathbf{y}[t] = \mathbf{H}\mathbf{x}[t]+\mathbf{A}\mathbf{s}[t]+\mathbf{z}[t],
\end{align}
where $\mathbf{x}[t]$ and $\mathbf{y}[t]$ represent channel input and output respectively of a vector channel at time index $t$, and $\mathbf{s}[t]$ and $\mathbf{z}[t]$ are zero-mean Gaussian random vectors i.i.d. across time with covariance matrices $\mathbf{Q}_s$, assumed to be full rank, and $\mathbf{Q}_z=I$ respectively. The dimension of $\mathbf{x}$, $\mathbf{s}$, $\mathbf{z}$, and $\mathbf{y}$ are $M_t$, $M_s$, $M_r$, and $M_r$ respectively. An $M_t \times M_r$ matrix $\mathbf{H}$ and an $M_s \times M_r$ matrix $\mathbf{A}$ are both assumed to be quasistatic in the sense that for any length-$n$ codeword they are constant. We use the exponent $n$, e.g., $\mathbf{x}^n\triangleq (\mathbf{x}[1],\mathbf{x}[2],\ldots,\mathbf{x}[n])$, to denote $n$ uses of the channel. On the input covariance matrix $\mathbf{Q}_x \triangleq \E\left[\mathbf{x}\mathbf{x}^\dagger\right]$, we impose the average power constraint $\tr(\mathbf{Q}_x)\leq P$.    

The transmitter is given the additive vector state (interference) sequence $\mathbf{s}^n$ noncausally, but knows only that $\mathbf{A}$ lies within an uncertainty set, $\mathbf{A}\in\mathcal{A}\subseteq\mathbb{R}^{M_s \times M_r}$ ($\mathbb{C}^{M_s \times M_r}$ for the complex channel). The uncertainty set $\mathcal{A}$ is defined as the set of all matrices with largest singular value bounded above by a known maximum amplification parameter $a_\mathsf{max}\in[0,\infty]$. Notice that $a_\mathsf{max}=\infty$ implies $\mathcal{A}=\mathbb{R}^{M_s \times M_r}$. Furthermore, the set $\mathcal{A}$ as defined is symmetric ($\mathbf{A}\in\mathcal{A}$ implies $-\mathbf{A}\in\mathcal{A}$), convex, and compact. 
When  $|\mathcal{A}|=1$, we have exactly the canonical vector dirty paper channel. 

Our analyses apply to both real- and complex-valued channels, and we note differences in assuming one or the other only as needed. For notation, we use boldfaced lowercase to represent vectors, boldfaced uppercase for matrices, $\dagger$ to denote Hermitian transpose, and all logarithms are base-2.



\section{Upper Bound on Capacity}
\label{sec:result}
\begin{theorem}[Upper Bound]
Define $M_0\triangleq\rank(\mathbf{H}\mathbf{Q}_x\mathbf{H}^\dagger)$ and a matrix $\mathbf{U}$ that projects the received signal onto the subspace spanned by $\mathbf{H}\mathbf{Q}_x\mathbf{H}^\dagger$. The capacity, $C$, is bounded above by $\overline{C}$ given in (\ref{eq:UB}),
\begin{figure*}[!t]
\normalsize
    \begin{align}
        \overline{C} \triangleq
            \sup_{\mathbf{Q}_x}\inf_{\{\mathbf{A}_i\}} &
            \frac{\kappa
            \left[{\displaystyle\sum_{i=1}^{\left\lceil\frac{M_s}{M_0}\right\rceil-1}}
            \log\frac{\det\left(\mathbf{U}\mathbf{H}\mathbf{Q}_x\mathbf{H}^\dagger \mathbf{U}^\dagger + \mathbf{I} + \mathbf{U}\mathbf{A}_i\mathbf{Q}_s\mathbf{A}_i^\dagger \mathbf{U}^\dagger\right)}{\det\left(\mathbf{U}\mathbf{A}_i\mathbf{Q}_s\mathbf{A}_i^\dagger \mathbf{U}^\dagger\right)}
            +\log\det\left(\mathbf{I}+\mathbf{U}\mathbf{H}\mathbf{Q}_x\mathbf{H}^\dagger \mathbf{U}^\dagger\right) + g\left(\mathbf{Q}_x,\mathbf{A}_{\left\lceil\frac{M_s}{M_0}\right\rceil}\right)\right]
            }{\left\lceil\frac{M_s}{M_0}\right\rceil+1}\label{eq:UB}\\
        g\left(\mathbf{Q}_x,\mathbf{A}\right)={}& 
            \begin{cases}
            \log\frac{\det\left(\mathbf{U}\mathbf{H}\mathbf{Q}_x\mathbf{H}^\dagger \mathbf{U}^\dagger + \mathbf{I} + \mathbf{U}\mathbf{A}\mathbf{Q}_s\mathbf{A}^\dagger \mathbf{U}^\dagger\right)}{\det\left(\mathbf{U}\mathbf{A}\mathbf{Q}_s\mathbf{A}^\dagger \mathbf{U}^\dagger\right)}
                & \text{ if }\left\lceil\frac{M_s}{M_0}\right\rceil = \frac{M_s}{M_0}\cr
            \log\frac{\det\left(\mathbf{U}\mathbf{H}\mathbf{Q}_x\mathbf{H}^\dagger \mathbf{U}^\dagger + \mathbf{I} + \mathbf{U}\mathbf{A}\mathbf{Q}_s\mathbf{A}^\dagger \mathbf{U}^\dagger\right)}{\det\left(\mathbf{U}\mathbf{A}\mathbf{Q}_s\mathbf{A}^\dagger \mathbf{U}^\dagger + \frac{1}{2}\mathbf{I}\right)} 
                + 2M_0 & \text{ else }
            \end{cases}.
\end{align}
\hrulefill
\vspace*{-15pt}
\end{figure*}
where $\kappa = \frac{1}{2}$ for real channels and $\kappa=1$ for complex channels, and supremum and infimum are subject to the constraints
    \begin{align}
        \mathbf{Q}_x\succeq{}& 0\ \nonumber\\
        \tr(\mathbf{Q}_x)\leq{}& P\ \nonumber\\
        \mathbf{A}_i\in{}&\mathcal{A}\ \forall\ i \in\left\{1,\ldots,\left\lceil\frac{M_s}{M_0}\right\rceil\right\}\nonumber\\
        \mathbf{A}_i\mathbf{Q}_s\mathbf{A}_j^\dagger ={}& 0\ \forall\ i\neq j\nonumber.
    \end{align}
\end{theorem}
Before presenting the proof, we point out that the received signal dimension is a \emph{chosen} integer, which affects the denominator of (\ref{eq:UB}). Consequently, the objective function of (\ref{eq:UB}) may be discontinuous with respect to $\mathbf{Q}_x$, implying that convex methods may not suffice to solve (\ref{eq:UB}) and that the maximin formulation is not necessarily interchangeable with a minimax formulation~\cite{Fan:1950}. Discontinuities also necessitate the use of supremum and infimum operations.
\begin{IEEEproof}
The proof begins in a manner similar to \cite{KELW:2007} and \cite{GS:2007:dyspan}, however differs in emphasis of potential state gain values, and is extended to vector state through an inductive argument. We assume real-valued channels for the presentation below.

We first fix the choice of $\mathbf{Q}_x$ and notice that at most $\left\lfloor\frac{M_s}{M_0}\right\rfloor$ rank-$M_0$ matrices, $\{\mathbf{A}_i\}_{i=1,\ldots,\left\lfloor\frac{M_s}{M_0}\right\rfloor}$, may be chosen such that $\mathbf{A}_i\mathbf{Q}_s\mathbf{A}_j^\dagger=0$. If $M_s$ is not evenly divisible by $M_0$, to collection $\{\mathbf{A}_i\}_{i=1,\ldots,\left\lfloor\frac{M_s}{M_0}\right\rfloor}$ we add a final matrix $\mathbf{A}_{\left\lceil\frac{M_s}{M_0}\right\rceil}$, which incorporates the remaining independent dimensions of $\mathbf{s}$. 
In the following, we denote the message as $W$, the channel output given state transformation matrix $\mathbf{A}$ as $\mathbf{y}_{\mathbf{A}}$, and the projected channel output as $\mathbf{v}_{\mathbf{A}}\triangleq\mathbf{U}\mathbf{y}_{\mathbf{A}}$. Noting that $\mathbf{A}_i\in\mathcal{A}$ implies $-\mathbf{A}_i\in\mathcal{A}$, we begin from Fano's inequality:
\begin{align}
nr  \leq{}& \min_{\mathbf{A}\in\mathcal{A}}I(W;\mathbf{y}_\mathbf{A}^n)\nonumber\\
    \stackrel{(a)}{\leq}{}& \min_{\mathbf{A}\in\mathcal{A}^\prime}I(W;\mathbf{v}_\mathbf{A}^n)\nonumber\\
    \stackrel{(b)}{\leq}{}& \frac{2I(W;\mathbf{v}_{\mathbf{0}}^n) 
        + \displaystyle\sum_{i=1}^{\left\lceil\frac{M_s}{M_0}\right\rceil}I(W;\mathbf{v}_{-\mathbf{A}_i}^n) 
            + I(W;\mathbf{v}_{\mathbf{A}_i}^n)}{2\left\lceil\frac{M_s}{M_0}\right\rceil+2}\nonumber\\
    ={}& \frac{2h(\mathbf{v}_{\mathbf{0}}^n) 
        + \displaystyle\sum_{i=1}^{\left\lceil\frac{M_s}{M_0}\right\rceil}h(\mathbf{v}_{-\mathbf{A}_i}^n) 
            + h(\mathbf{v}_{\mathbf{A}_i}^n)}{2\left\lceil\frac{M_s}{M_0}\right\rceil+2}\nonumber\\
    &- \frac{2h(\mathbf{v}_{\mathbf{0}}^n|W) 
        + \displaystyle\sum_{i=1}^{\left\lceil\frac{M_s}{M_0}\right\rceil}h(\mathbf{v}_{-\mathbf{A}_i}^n|W) 
            + h(\mathbf{v}_{\mathbf{A}_i}^n|W)}{2\left\lceil\frac{M_s}{M_0}\right\rceil+2},\label{eq:ob0}
\end{align}
where in step (a) we chose a reduced uncertainty set
\begin{align*}
\mathcal{A}^\prime = \left\{-\mathbf{A}_{\left\lceil\frac{M_s}{M_0}\right\rceil},\ldots,-\mathbf{A}_1,\mathbf{0},\mathbf{A}_1,\ldots,\mathbf{A}_{\left\lceil\frac{M_s}{M_0}\right\rceil}\right\},
\end{align*}
and project onto the subspace containing the message, and in (b) we note that the arithmetic mean is greater than the minimum of a set.

First, we bound the unconditioned entropy terms of antipodal state channel matrices:
\begin{align}
h(&\mathbf{v}_{-\mathbf{A}_i}^n) + h(\mathbf{v}_{\mathbf{A}_i}^n)\nonumber\\
    \stackrel{(a)}{\leq}{}& \sum_{t=1}^{n}\frac{1}{2}\log\det\left(\mathbf{I}+\mathbf{U}\mathbf{H}\mathbf{Q}_{x[t]}\mathbf{H}^\dagger \mathbf{U}^\dagger + \Psi_t + \mathbf{U}\mathbf{A}_i\mathbf{Q}_s\mathbf{A}_i^\dagger \mathbf{U}^\dagger\right)\nonumber\\
& + \frac{1}{2}\log\det\left(\mathbf{I}+\mathbf{U}\mathbf{H}\mathbf{Q}_{x[t]}\mathbf{H}^\dagger \mathbf{U}^\dagger - \Psi_t + \mathbf{U}\mathbf{A}_i\mathbf{Q}_s\mathbf{A}_i^\dagger \mathbf{U}^\dagger\right)\nonumber\\
& + nM_0\log(2\pi e)\nonumber\\
    \stackrel{(b)}{\leq}& n\log\det\left(\mathbf{I}+\mathbf{U}\mathbf{H}\mathbf{Q}_{x}\mathbf{H}^\dagger \mathbf{U}^\dagger + \mathbf{U}\mathbf{A}_i\mathbf{Q}_s\mathbf{A}_i^\dagger \mathbf{U}^\dagger\right)\nonumber\\
& + nM_0\log(2\pi e),\label{eq:obA}
\end{align}
where (a) uses maximum entropy principles and expansion of covariance terms, (b) uses the concavity of log-determinant, and
$Q_{x[t]}$ and $\Psi_t\triangleq\E[\mathbf{U}\mathbf{H}\mathbf{x}[t]\mathbf{s}^\dagger \mathbf{A}_i^\dagger\mathbf{U}^\dagger + \mathbf{U}\mathbf{A}_i\mathbf{s}\mathbf{x}[t]^\dagger \mathbf{H}^\dagger\mathbf{U}^\dagger]$ denote input covariance and cross correlation of $t$-th channel use respectively.
Additionally, we note
\begin{align}
h(\mathbf{v}_{\mathbf{0}}^n) \leq{}& \frac{n}{2}\log\det\left(\mathbf{I}+\mathbf{U}\mathbf{H}\mathbf{Q}_x\mathbf{H}^\dagger \mathbf{U}^\dagger\right) + \frac{nM_0}{2}\log(2\pi e).\label{eq:obB}
\end{align}

We lower bound terms conditioned on the message $W$:
\begin{align}
h(\mathbf{v}_{\mathbf{0}}^n&|W) + \displaystyle\sum_{i=1}^{\left\lceil\frac{M_s}{M_0}\right\rceil}h(\mathbf{v}_{\mathbf{A}_i}^n|W)\nonumber\\
    \geq{}& h(\mathbf{v}_{\mathbf{0}}^n,\mathbf{v}_{\mathbf{A}_1}^n,\ldots,\mathbf{v}_{\mathbf{A}_{\left\lceil\frac{M_s}{M_0}\right\rceil}}^n|W)\nonumber\\
    \stackrel{(a)}{=}{}& h\left(\frac{\mathbf{v}_{\mathbf{A}_1}^n+\mathbf{v}_{\mathbf{0}}^n}{\sqrt{2}},\frac{\mathbf{v}_{\mathbf{A}_1}^n-\mathbf{v}_{\mathbf{0}}^n}{\sqrt{2}},\mathbf{v}_{\mathbf{A}_2}^n,\ldots,\mathbf{v}_{\mathbf{A}_{\left\lceil\frac{M_s}{M_0}\right\rceil}}^n\middle|W\right)\nonumber\\
    ={}& h\left(
        \frac{\mathbf{v}_{\mathbf{A}_1}^n+\mathbf{v}_{\mathbf{0}}^n}{\sqrt{2}},
        \mathbf{v}_{\mathbf{A}_2}^n,\ldots,\mathbf{v}_{\mathbf{A}_{\left\lceil\frac{M_s}{M_0}\right\rceil}}^n 
        \middle| \frac{\mathbf{v}_{\mathbf{A}_1}^n-\mathbf{v}_{\mathbf{0}}^n}{\sqrt{2}},W \right)\nonumber\\
    & + h\left(\frac{\mathbf{v}_{\mathbf{A}_1}^n-\mathbf{v}_{\mathbf{0}}^n}{\sqrt{2}}|W\right)\nonumber\\
    \stackrel{(b)}{=}{}& h\left(\frac{\mathbf{v}_{\mathbf{A}_1}^n+\mathbf{v}_{\mathbf{0}}^n}{\sqrt{2}},\mathbf{v}_{\mathbf{A}_2}^n,\ldots,\mathbf{v}_{\mathbf{A}_{\left\lceil\frac{M_s}{M_0}\right\rceil}}^n\middle|\frac{\mathbf{U}\mathbf{A}_1 \mathbf{s}^n}{\sqrt{2}},W\right)\nonumber\\
    & + h\left(\frac{\mathbf{U}\mathbf{A}_1 \mathbf{s}^n}{\sqrt{2}}\right)\nonumber\\
%
    \stackrel{(c)}{\geq}{}& h(\mathbf{v}_{\mathbf{0}}^n,\mathbf{v}_{\mathbf{A}_2}^n,\ldots,\mathbf{v}_{\mathbf{A}_{\left\lceil\frac{M_s}{M_0}\right\rceil}}^n|W)\nonumber\\
    & + \frac{n}{2}\log\det\left(\mathbf{U}\mathbf{A}_1 \mathbf{Q}_s \mathbf{A}_1^\dagger \mathbf{U}^\dagger\right) + \frac{nM_0}{2}\log(2\pi e),\label{eq:ob3}
\end{align}
where (a) results from a basis transformation, (b) results from perfectly correlated noise terms between the two channel outputs and independence between message and state, and (c) results from factorization of matched scaling constants. The analysis for (\ref{eq:ob3}) is repeated inductively to arrive at
\begin{align}
h(\mathbf{v}_{\mathbf{0}}^n&|W) + \displaystyle\sum_{i=1}^{\left\lceil\frac{M_s}{M_0}\right\rceil}h(\mathbf{v}_{\mathbf{A}_i}^n|W)\nonumber\\
\geq{}&h(\mathbf{v}_{\mathbf{0}}^n,\mathbf{v}_{\mathbf{A}_{\left\lceil\frac{M_s}{M_0}\right\rceil}}^n|W)
+ \frac{nM_0\left(\left\lceil\frac{M_s}{M_0}\right\rceil-1\right)}{2}\log(2\pi e)\nonumber\\ 
&  + \sum_{i=1}^{\left\lceil\frac{M_s}{M_0}\right\rceil-1} \frac{n}{2}\log\det\left(\mathbf{U}\mathbf{A}_i \mathbf{Q}_s \mathbf{A}_i^\dagger \mathbf{U}^\dagger\right).\nonumber
\end{align}

If $M_s$ is evenly divisible by $M_0$ the same induction may be applied to decouple the final two potential channel outputs:
\begin{align}
    h(\mathbf{v}_{\mathbf{0}}^n,&\mathbf{v}_{\mathbf{A}_{\left\lceil\frac{M_s}{M_0}\right\rceil}}^n|W) \nonumber\\
        \geq{}& \frac{nM_0}{2}\log(2\pi e) + \frac{n}{2}\log\det\left(\mathbf{U}\mathbf{A}_{\left\lceil\frac{M_s}{M_0}\right\rceil} \mathbf{Q}_s \mathbf{A}_{\left\lceil\frac{M_s}{M_0}\right\rceil}^\dagger \mathbf{U}^\dagger\right).\label{eq:obC}
\end{align}

If $M_s$ is not evenly divisible by $M_0$, the matrix $\mathbf{U}\mathbf{A}_{\left\lceil\frac{M_s}{M_0}\right\rceil}\mathbf{Q}_s\mathbf{A}_{\left\lceil\frac{M_s}{M_0}\right\rceil}^\dagger \mathbf{U}^\dagger$ is rank deficient, and thus we only partially correlate the two noise terms $\mathbf{z}_{\left\lceil\frac{M_s}{M_0}\right\rceil}$ and $\mathbf{z}_{0}^n$:
\begin{align}
    h(&\mathbf{v}_{\mathbf{0}}^n,\mathbf{v}_{\mathbf{A}_{\left\lceil\frac{M_s}{M_0}\right\rceil}}^n|W) \nonumber\\
    ={}& h\left(
        \frac{\mathbf{v}_{\mathbf{A}_{\left\lceil\frac{M_s}{M_0}\right\rceil}}^n+\mathbf{v}_{\mathbf{0}}^n}{\sqrt{2}}
        \middle| \frac{\mathbf{U}}{\sqrt{2}}\left(\mathbf{A}_{\left\lceil\frac{M_s}{M_0}\right\rceil}\mathbf{s}^n + \mathbf{z}_{\left\lceil\frac{M_s}{M_0}\right\rceil}-\mathbf{z}_{0}^n\right),
         W \right)\nonumber\\
    & + h\left(\frac{\mathbf{U}}{\sqrt{2}}\left(\mathbf{A}_{\left\lceil\frac{M_s}{M_0}\right\rceil}\mathbf{s}^n + \mathbf{z}_{\left\lceil\frac{M_s}{M_0}\right\rceil}-\mathbf{z}_{0}^n\right)\middle|W\right)\nonumber\\
    \geq{}& h\left(
        \frac{\mathbf{U}}{\sqrt{2}}\left(\mathbf{z}_{\left\lceil\frac{M_s}{M_0}\right\rceil}+\mathbf{z}_{0}^n\right)\right)\nonumber\\
        & + h\left(\frac{\mathbf{U}}{\sqrt{2}}\left(\mathbf{A}_{\left\lceil\frac{M_s}{M_0}\right\rceil}\mathbf{s}^n + \mathbf{z}_{\left\lceil\frac{M_s}{M_0}\right\rceil}-\mathbf{z}_{0}^n\right)\middle|W\right)\nonumber\\
     \stackrel{}{\geq}{}& \frac{nM_0}{2}\log\left(\pi e\right) + \frac{n}{2}\log\det\left(\mathbf{U}\mathbf{A}_{\left\lceil\frac{M_s}{M_0}\right\rceil} \mathbf{Q}_s \mathbf{A}_{\left\lceil\frac{M_s}{M_0}\right\rceil}^\dagger \mathbf{U}^\dagger+\frac{1}{2}\mathbf{I}\right).\label{eq:noisecorr2}
\end{align}

An identical analysis may be performed for $h(\mathbf{v}_{\mathbf{0}}^n|W) + \sum_{i=1}^{\left\lceil\frac{M_s}{M_0}\right\rceil}h(\mathbf{v}_{-\mathbf{A}_i}^n|W)$,
and by substituting (\ref{eq:obA}), (\ref{eq:obB}), (\ref{eq:obC}), and (\ref{eq:noisecorr2}) for both positive and negative $\mathbf{A}_i$ into (\ref{eq:ob0}) and allowing minimization over collections of $\{\mathbf{A}_i\}$, we arrive at (\ref{eq:UB}).\end{IEEEproof}

\noindent\textbf{Remark 1:} Although the optimization in (\ref{eq:UB}) is potentially difficult to compute, a simpler bound can be arrived at which chooses as each $\mathbf{A}_i$ a matrix that aligns eigenvectors of $\mathbf{Q}_s$ with eigenvectors of $\mathbf{H}\mathbf{Q}_x\mathbf{H}^\dagger$.

\noindent\textbf{Remark 2:} For the case where both the channel input and interference are scalars, the bound of \cite{GS:2007:dyspan} provided evidence that, with unknown phase, correlation between the input and state provided no benefit, and comparison between a zero additive state with a high-variance additive state is a special case studied in \cite{KELW:2007}. Incorporating both of these emphases into a single analysis provides new insight into how the unknown phase and unknown amplitude jointly reduce capacity.

\noindent\textbf{Remark 3:} The primary innovation in the construction of (\ref{eq:UB}) is emphasis of the effect that the dimension of interference may have on the prelog factor of capacity. In particular, if large but finite interference power is considered, i.e., $P \ll a_\mathsf{max}^2<\infty$, the terms in the sum of (\ref{eq:UB}) tend towards zero signifying an approximate prelog capacity loss. The error in this approximation depends primarily on the ${\rm SIR}$ of the system. If $a_\mathsf{max}=\infty$, i.e. the set $\mathcal{A}=\mathbb{R}^{M_s\times M_r}$, then the prelog loss becomes exact and the system exhibits what resembles a degrees of freedom loss \emph{at all ${\rm SNR}$}.

\noindent\textbf{Remark 4:} The choice of correlation of noise terms in (\ref{eq:noisecorr2}) is not optimized, and thus the bound is potentially loose. This optimization however depends on $\mathbf{Q}_s$ and choice of $\{\mathbf{A}_i\}$ which in turn depends on choice of $\mathbf{Q}_x$. As discussed prior to presentation of the proof, it not immediately apparent how these choices interact to tighten or loosen the bound.

\section{MISO \& SIMO Channels with Vector State}
\label{sec:rankone}

The question of the dimension of the received message bearing signal $\mathbf{H}\mathbf{x}$, and subsequent optimization of the input covariance $\mathbf{Q}_x$ prevents a more explicit characterization of (\ref{eq:UB}) in general. On the other hand, if the signal is necessarily one-dimensional (e.g., when either the transmitter or receiver in a wireless communication link has one antenna) the upper bound on capacity may be simplified considerably:
\begin{corollary}
    Let $v_{si}$ denote the eigenvalues of $\mathbf{Q}_s$, and $\mathbf{h}$ the channel vector modifying $\mathbf{x}$. The capacity, $C$, is bounded above by $\overline{C}_1$ given by
    \begin{align}
        \overline{C}_1 =& \frac{\kappa \left[{\displaystyle\sum_{i=1}^{M_s}}\log\left(\frac{\|\mathbf{h}\|^2P+1+a_\mathsf{max}^2v_{si}}{a_\mathsf{max}^2v_{si}}\right)
        +\log\left(1+\|\mathbf{h}\|^2P\right)\right]}{M_s+1},\label{eq:corr}
    \end{align}
    where $\kappa = \frac{1}{2}$ for real channels and $\kappa=1$ for complex channels.
\end{corollary}
\begin{sketch}
    The full proof is omitted due to limited space, however it relies only on beamforming (transmit or receive) for the message-bearing signal and a sequence of $\mathbf{A}_i$ matrices that project individual eigenvectors of $\mathbf{Q}_s$ onto the subspace containing the message-bearing signal.
\end{sketch}

\noindent \textbf{Remark 5:} The point made in Remark~3 regarding approximation of the bound with a prelog factor is more clearly illustrated in rank-1 channels. If for example 
\begin{align}
    \min_i v_{si}\geq\frac{1+\|\mathbf{h}\|^2P}{a_\mathsf{max}^2},
\end{align}
then each log term in the sum of (\ref{eq:corr}) is bounded above by 1, and the gap between the approximation ${\widetilde{C}_1\triangleq \frac{\kappa}{M_s+1}\log(1+\|\mathbf{h}\|^2P)}$ and the actual bound is less than $\frac{\kappa M_s}{M_s+1}$ bits.

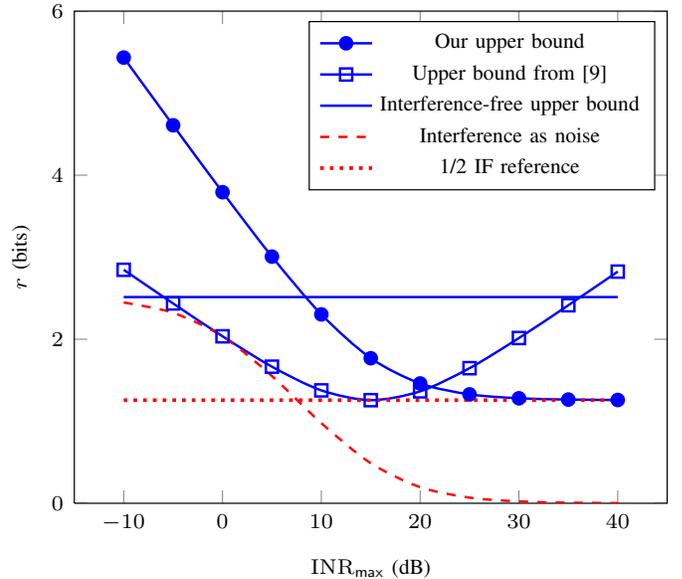
\begin{figure}[t]
\centering
\begin{tikzpicture}[scale=1.15,font=\scriptsize,%
every axis/.style={
    ymax=6,%
    ymin=0,%
    ytick={0,2,...,6}}]
    \begin{axis}[
        xlabel=${\rm INR}_\mathsf{max}$ (dB),
        ylabel=$r$ (bits),
        every axis y label/.style=
            {at={(ticklabel cs:0.5)},rotate=90,anchor=near ticklabel}]
    \addplot[smooth,thick,mark=*,blue] plot file {ks.data};
    \addlegendentry{Our upper bound}
    \addplot[smooth,thick,color=blue,mark=square]
        plot file {gs.data};
    \addlegendentry{Upper bound from \cite{GS:2007:dyspan}}
    \addplot[smooth,thick,color=blue]
        plot coordinates {
            (-10,2.5139)
            (40,2.5139)
        };
    \addlegendentry{Interference-free upper bound}
    \addplot[smooth,thick,color=red,dashed]
        plot file {tin.data};
    \addlegendentry{Interference as noise}
    \addplot[smooth,very thick,color=red,dotted]
        plot coordinates {
            (-10,1.257)
            (40,1.257)
        };
    \addlegendentry{1/2 IF reference}
    \end{axis}
    \end{tikzpicture}
\caption{Comparison of upper and lower bounds for the compound dirty paper channel with real scalar input and interference (${\rm SNR}=$ 15 dB). The dotted trace represents half the interference-free rate and is included as reference; it does not represent an achievable scheme.}\label{fig:cap}\vspace{0pt}
\end{figure}

\noindent \textbf{Remark 6:} One special case is the scalar channel with scalar additive state whose bound is shown in Figure~\ref{fig:cap} relative to prior work. Notice our bound is tighter at high ${\rm INR}$ (low ${\rm SIR}$), and complements the result from~\cite{GS:2007:dyspan}. The approximate prelog loss is illustrated as well: for the scalar case, $M_s = 1$ so the prelog loss is approximately $\frac{1}{2}$.

\section{High-${\rm SNR}$ Behavior}
\label{sec:highSNR}

The behavior of wireless systems at high signal to noise ratios often provides insight into the spatial interaction of signals. One standard metric for high-${\rm SNR}$ performance is the multiplexing gain or degrees of freedom ($\mathsf{DOF}$) defined as
\begin{align}
    \mathsf{DOF} \triangleq \lim_{{\rm SNR}\rightarrow\infty} \frac{C({\rm SNR})}{\kappa\log(1+{\rm SNR})},
\end{align}
where ${\rm SNR}=P$, and $\kappa = \frac{1}{2}$ or $\kappa=1$ for real or complex channels respectively. In our system, an upper bound on $\mathsf{DOF}$ may be deduced directly from (\ref{eq:UB}). The form of the bound is contingent on how the ${\rm INR}$, or equivalently the spectral properties of state sequence $\mathbf{s}^n$, scales with ${\rm SNR}$:\vspace{0.5em}
\begin{corollary}
The system has full ($M^\star\triangleq\min\left\{M_t,M_r\right\}$) degrees of freedom if and only if both of the following conditions hold:
\begin{enumerate}
    \item The parameter $a_\text{max}$ is finite.
    \item The interference power, or equivalently ${\rm INR}$, grows sublinearly with respect to ${\rm SNR}$.
\end{enumerate}
If either condition does not hold then the degrees of freedom of the system is bounded above by
    \begin{align}
        \mathsf{DOF} \leq{}& \frac{M^\star\left(\left\lceil\frac{M_s}{M^\star}\right\rceil+1\right) - M_s}{\left\lceil\frac{M_s}{M^\star}\right\rceil+1}.\label{eq:UBDOF}
    \end{align}\label{corr:DOF}
\end{corollary}
\begin{IEEEproof}
If both conditions hold, then the interference may be treated as noise and at high ${\rm SNR}$ the gap between the rate achieved and the interference free rate approaches a constant. On the other hand, if the first condition is false, then, as noted in Remark~2, the terms in the summation of (\ref{eq:UB}) evaluate to 0. Therefore this proof focuses on the case where only the second condition is false. 

If ${\rm INR}$ grows linearly with ${\rm SNR}$, then there must exist some finite scalar $\beta$ such that
\begin{align}
   \mathbf{H}\mathbf{Q}_x\mathbf{H}^\dagger\preceq \beta \mathbf{A}\mathbf{Q}_s\mathbf{A}^\dagger,\label{eq:linbound}
\end{align}
for all $\mathbf{Q}_x$ where $\tr(\mathbf{Q}_x)\leq P$. Consequently, the terms in the summation of (\ref{eq:UB}) may be bounded by a constant which has vanishing contribution when normalized by $\log(1+P)$ as $P\rightarrow\infty$. 
If ${\rm INR}$ grows superlinearly with respect to ${\rm SNR}$, then a function $\beta(P)\rightarrow 0$ as $P\rightarrow\infty$ suffices to satisfy (\ref{eq:linbound}), and the the terms in the summation of (\ref{eq:UB}) vanish as $P\rightarrow\infty$. 

By counting the number of dimensions of interference relative to message bearing signal, the asymptotic behavior of the remaining two terms in the numerator of (\ref{eq:UB}) results in the $\mathsf{DOF}$ upper bound for fixed $M_0$
\begin{align}
        \mathsf{DOF} \leq{}& \frac{M_0\left(\left\lceil\frac{M_s}{M_0}\right\rceil+1\right) - M_s}{\left\lceil\frac{M_s}{M_0}\right\rceil+1},\nonumber
\end{align}
which is maximized when $M_0 = M^\star$.
\end{IEEEproof}

\noindent \textbf{Remark 7:} 
It is important to note cases where the two conditions posed in Corollary~\ref{corr:DOF} hold in the context of common wireless network applications of DPC. With respect to the first condition, often the known sequence represents encoded messages intended for other receivers that interfere with the DPC transmission. In these cases, perhaps when the ${\rm INR}$ is high enough, i.e., the singular values of $\mathbf{A}$ are above $a_\mathsf{max}$, the interference may be decoded and the nature of the system changes. Alternatively, the vector $\mathbf{s}$ might represent multiple known sequences whose exact linear transformation at the receiver is unknown, but whose magnitude may be bounded based on a measurement of aggregate ${\rm INR}$. 

For the second condition to hold, we must assume that the nature of increased ${\rm SNR}$ is a result of increased transmssion power at the transmitter rather than a decrease in thermal noise at the receiver.

\noindent \textbf{Remark 8:} Unlike the finite-${\rm SNR}$ behavior, the $\mathsf{DOF}$ loss exhibited when Condition~2 is false is not confined to any specific ${\rm SIR}$ regime. Even if $a_\mathsf{max}$ is small, if the ${\rm INR}$ scales linearly with ${\rm SNR}$ then the statement holds. Moreover, the exact covariance structure is less relevant at high-${\rm SNR}$ than the dimension or rank of $\mathbf{Q}_s$.


\section{Summary}
\label{sec:remarks}

In this paper, we studied a compound channel model for vector dirty paper where the linear transformation spinning and stretching the dirty paper is unknown. We presented an upper bound on the capacity of this compound vector dirty paper channel. Our bound is tighter than previous bounds in the low-${\rm SIR}$ regime for the case of scalar input and interference, and extends intuitions regarding prelog loss in capacity to the vector dirty paper channel. The bound offers insight into the high-${\rm SNR}$ behavior of systems modeled by vector dirty paper, and a relationship between the dimension of the message-bearing signal, the dimension of the interference, and the degrees of freedom of the system was revealed.

\bibliographystyle{IEEEtran}
\bibliography{references}

\end{document}